\documentclass{IEEEtran}
\usepackage{cite}
\usepackage{amsmath,amssymb,amsfonts}
\usepackage{algorithm}
\usepackage{graphicx}
\usepackage{textcomp}
\usepackage{comment,color}
\usepackage{algpseudocode}

\newcommand{\R}{\ensuremath{{\mathbb R}}}
\newcommand{\Z}{\ensuremath{{\mathbb Z}}}

\newcommand{\uuu}{{\mathsf{u}}}

\newcommand{\nnn}{{\mathsf{n}}}
\newcommand{\mmm}{{\mathsf{m}}}
\newcommand{\ppp}{{\mathsf{p}}}

\newcommand{\rrr}{{\mathsf{r}}}
\newcommand{\sss}{{\mathsf{s}}}

\newcommand{\ol}{\overline}

\newcommand{\UUUU}{{\mathcal U}}

\newtheorem{thm1}{\bf Theorem}
\newtheorem{prop1}{\bf Proposition}
\newtheorem{lem1}{\bf Lemma}
\newtheorem{asm1}{\bf Assumption}
\newtheorem{defn1}{\bf Definition}
\newtheorem{rem1}{\bf Remark}
\newtheorem{cor1}{\bf Corollary}

\newenvironment{rem}{\begin{rem1}}{\hfill$\square$\end{rem1}}

\newenvironment{cor}{\begin{cor1}}{\hfill$\square$\end{cor1}}
\newenvironment{prop}{\begin{prop1}}{\hfill$\square$\end{prop1}}

\def\BibTeX{{\rm B\kern-.05em{\sc i\kern-.025em b}\kern-.08em
    T\kern-.1667em\lower.7ex\hbox{E}\kern-.125emX}}
\begin{document}
\title{Further Methods for Encrypted Linear Dynamic Controllers Utilizing Re-Encryption}
\author{Junsoo Kim, \IEEEmembership{Member, IEEE}
\thanks{This paper is supported by Seoul National University of Science and Technology.}
\thanks{The author is with the Department of Electrical and Information Engineering, Seoul National University of Science and Tecchnology, Korea (e-mail: junsookim@seoultech.ac.kr).}
}

\maketitle

\begin{abstract}
Homomorphic encryption (HE) applied to a networked controller enables secure operation, but in most cases it allows for addition and multiplication over integers only, because of computation efficiency. Several related results deal with such constraints by means of re-encrypted controller output, based on which the controller can be re-constructed and operate over integers. This paper presents two modified methods based on the output re-encryption, which will further reduce the required communication effort and computational burden, respectively.
\end{abstract}


\section{Introduction}
Encrypted control is to employ recent cryptosystems to protect all data in the network layer by encryption, which allow for computation directly over encrypted data without decryption \cite{Kogiso15CDC,SchulzeDarup21CSM,Kim22ARC}.
Not only the communication stage but also the computation stage can be protected by encryption contrary to conventional encryption,
and it does not sacrifice precision for the purpose of security, contrary to, such as, differential privacy methods.

Methods based on homomorphic encryption (HE) solely,
which exploit computationally less expensive addition and multiplication,
have a competitive advantage compared to bootstrapping or multi-party computation based methods, in terms of efficiency or security, respectively \cite{Kim22ARC}.
However, there is a {\it constraint on the operation} as a trade-off; once encrypted and transmitted to the computation unit, {\it functions other than addition and multiplication (over integers) are limited}.
As a consequence, systems to apply HE need to be re-formulated to fit in this constraint.

For example, \cite{Kogiso15CDC} considered how to dealt with limitation of addition, when applying multiplicatively HE to dynamic controllers.
Periodic reset has been suggested in \cite{Murguia20TAC}, to refresh the encrypted state of system to reuse for the operation.
To address incapability of recursive multiplication by non-integer numbers, \cite{Kim23TAC} introduced a concept of converting state matrix to integers.
And, it was followed by \cite{TeranishiTCNS,LeeTSMC} in which auto-regressive model representation methods are presented, to deal with further limitations with use of HE.
Stability of such dynamic systems having state matrix as integers has been studied in \cite{SchulzeDarup22TAC}.
Furthermore, the limitation on operation with HE has also been considered and handled; in data-driven methods \cite{AlexandruCDC20,Alisic23CSL}, protocols for privacy in multi-agent systems \cite{Wang19TAC,Hadjicostis20TAC,Lee20CDC}, discrete event systems \cite{Genise19ASIACRYPT,Fritz19ACC}, and quantization methods for controllers \cite{Jayawardhana23CSL}.

\subsection{Methods Using Output Re-Encryption}

To address this constraint with HE applied in networked control systems,
several results have presented methods for re-formulating dynamic systems to operate over integers, using ``re-encryption of output'' \cite{Kim23TAC,TeranishiTCNS,LeeTSMC}.
Consider a dynamic controller written as
\begin{align}\label{eq:controller}
	\begin{split}
		x(t+1) &= Fx(t) + Gy(t),
		~~~~ x(0)=x_0,
		\\
		u(t) &= Hx(t),
	\end{split}	
\end{align}
in which $x(t)\in\R^\nnn$ is the state, $y(t)\in\R^\ppp$ is the input, and $u(t)\in\R^\mmm$ is the output of the system, respectively.
The problem is due to recursive multiplication by the state matrix $F\in\R^{\nnn\times\nnn}$ which consists of non-integer numbers in general, so that as time goes by ($t=0,1,\ldots$), it eventually results in necessity of rounding operation for discarding least significant figures. See \cite[Eq.~(32)]{Kim22ARC} or \cite[Section~II.D]{Kim23TAC} for more details.

The issue can be resolved by re-encryption of the controller output $u(t)$, which helps the state matrix $F$ to be converted to an integer matrix.
Let the system \eqref{eq:controller} be re-written as
\begin{subequations}\label{eq:reenc}
	\begin{align}
		x(t+1) &= (F-RH)x(t) + Gy(t)+Ru(t)\label{eq:reenc_a}\\
		u(t) &= Hx(t)\label{eq:reenc_b}
	\end{align}
\end{subequations}
where $R\in\R^{\nnn\times \mmm}$ is a matrix to be chosen. Note that the portion $RHx(t)$ for the state has been substituted by the term $Ru(t)$ which will be regarded as an auxiliary input of the system. Obviously, the choice of the matrix $R$ does not affect the input-output relation (from $y(t)$ to $u(t)$) or the performance of the system,
as long as the same output $u(t)$ of \eqref{eq:reenc_b} is fed back to the state in \eqref{eq:reenc_a} for each time.

Then,
the result of \cite{Kim23TAC} is that
the matrix $R$ can be appropriately chosen so that $F-RH$ can be transformed to an integer matrix. And, with $F-RH$ transformed to integers, the system \eqref{eq:reenc} can continue the operation over encrypted data exploiting addition and multiplication only, assuming that the portion $u(t)$ in \eqref{eq:reenc_a} is re-encrypted (this will be reviewed in the next sections).
The results in \cite{TeranishiTCNS} and \cite{LeeTSMC} further showed that the re-encryption also enables the controller to compute the state and the output, using only finite number of times of multiplication.
Although some follow-up results \cite{Kim21CDC,Tavazoei23TAC} offloaded the requirement of re-encryption, due to enlarged dimension or time-varying implementation, they cost more amount of controller storage or computational resource for implementation, compared to the methods with re-encryption.

\subsection{Contribution}
This brief paper presents two further methods for encrypting linear systems with use of output re-encryption, which will improve on efficiency of the controller design or the required communication effort.
Section~\ref{subsec:output}
will show that,
considering a chain of subspaces with respect to observability of the pair $(F,H)$,
the matrix $R$ can be chosen so that both the state matrix $F-RH$ and the output matrix $H$ can be transformed to integer matrices. This will reduce the required size of the plaintext space of the cryptosystem, which implies less use of storage and computational resource of encrypted controllers.
Contrary to \cite[Section~III.E]{Kim23TAC}, the simultaneous transformation to integers is also applicable for multi-output systems.

In addition to the state and output matrices converted to integers, another note regarding the input matrix is that it can be kept as rational numbers, in case the given control parameters $\{F,G,H\}$ are rational matrices.
This will imply that, after the controller conversion, the input matrix can be kept in digital computers without a truncation error, not to sacrifice the control performance.

In contrast to the previous results \cite{Kim23TAC,TeranishiTCNS,LeeTSMC} assuming
the re-encrypted signal provided
every sampling period,
Section~\ref{subsec:intermittent} will
show that
the controller can be implemented to operate over integers, even though the re-encryption is performed from time to time, intermittently.
This will imply that the encrypted dynamic operation can be persisted although the period of using decryption key for re-encryption is prolonged.

{\it Notation:} Let $\R$ and $\Z$ denote the set of real numbers and integers, respectively.
For matrices and vectors,  let $\lceil\cdot\rfloor$ denote the component-wise rounding operation,
and $I_\nnn\in\R^{\nnn\times\nnn}$ denote the identity matrix.
For a matrix $H\in\R^{\mmm\times \nnn}$, define $\ker H:=\{x\in\R^\nnn:Hx=0\}\subseteq\R^\nnn$.
If there is no ambiguity,
we abuse notation and let
$0$ also denote a zero matrix with appropriate dimensions.

\section{Main Result}
Recall that the problem is to implement the controller \eqref{eq:controller} or \eqref{eq:reenc} to operate over integers using $(+,\times)$, and to this end, the state matrix should be converted to integers.

\subsection{Method for Having Output Matrix as Integers}\label{subsec:output}

We show that, by appropriate choice of the matrix $R$ in \eqref{eq:reenc}, both the state matrix and the output matrix can be transformed to integer matrices. Let $T\in\R^{\nnn\times\nnn}$ be an invertible matrix, with which \eqref{eq:reenc} can be transformed as
\begin{subequations}\label{eq:controller_transformed}
	\begin{equation}
			z(t+1) = T(F-RH)T^{-1}z(t) + TGy(t)+TRu(t)
		\end{equation}
		where $z(t):= Tx(t)\in\R^\nnn$ is the transformed state with $z(0) = Tx_0$.
		We let the output be computed by
		\begin{equation}
			u_z(t) = T_uHT^{-1}z(t)\in\R^\mmm
		\end{equation}
		with some invertible matrix $T_u\in\R^{\mmm\times \mmm}$,
		from which the real output $u(t)$ can be restored by 
		$u(t) = T_u^{-1}u_z(t)$.
	\end{subequations}
	The objective is to choose $T$, $R$, and $T_u$ such that
	\begin{equation}\label{eq:output_integer}
		T(F-RH)T^{-1}\in\Z^{\nnn\times \nnn}~~ \text{and}~~ T_uHT^{-1}\in\Z^{\mmm\times\nnn}.
	\end{equation}

	Observable canonical form of the controller will be exploited for the design,
	so without loss of generality,
	we first assume that the controller \eqref{eq:controller} (i.e., the pair $(F,H)$) is observable.
	Indeed, if there is unobservable portion of the state that does not affect the output $u(t)$, the system can be reduced to be observable which will not change the input-output relation of the controller.
	And, we assume that the output matrix $H\in\R^{\mmm\times \nnn}$ is of full row rank (so $\mmm \le \nnn$), which is satisfied in most cases.
	
	To find $T$ and $R$,
	we consider a chain of subspaces, as
	\begin{equation}\label{eq:chain}
		\{0\}= \UUUU_{\nnn} \subseteq\UUUU_{\nnn-1}
		\subseteq \cdots\subseteq\UUUU_{1}
		\subseteq \UUUU_0 = \R^\nnn
	\end{equation}
		where
		\begin{equation}\label{eq:unobs}
			\UUUU_i := \ker \begin{bmatrix}
				H\\HF\\\vdots\\HF^{i-1}
			\end{bmatrix},\quad i= 1,2,\ldots, \nnn.
		\end{equation}
		Then, the following algorithm shows how the transformation $T$ is determined, by a basis representation in $\R^\nnn$.
		
\begin{algorithm}[h]
	\caption{Construction of transformation $T\in\R^{\nnn\times\nnn}$.}\label{alg1} 
	\begin{algorithmic}[1]
		\renewcommand{\algorithmicrequire}{\textbf{Input:}}
		\Require The matrices $F\in\R^{\nnn\times\nnn}$ and $H\in\R^{\mmm\times\nnn}$ from \eqref{eq:controller}.
		\renewcommand{\algorithmicensure}{\textbf{Setup:}}
		\Ensure Calculate the spaces $\{\UUUU_i\}_{i=0}^{\nnn}$.
		\State Find a matrix $V_1$ whose columns compose a basis of the space $\UUUU_{\nnn-1}$.
		\State Define $W_1:=V_1$.
		\For{$i=2,\ldots,\nnn-1$}
		\State Find a matrix $V_i$ such that the columns of
		$$\begin{bmatrix}
			W_1&W_2&\cdots& W_{i-1}& FW_{i-1}&V_i
		\end{bmatrix}$$
		
		are a basis of $\UUUU_{\nnn-i}$.
		\State Define $W_i:=[FW_{i-1},V_i]$.
		\EndFor
		\State Find $V_n$ such that the columns of
		$\{W_i\}_{i=1}^{\nnn-1}$ and $W_\nnn:=[FW_{\nnn-1},V_\nnn]$ are a basis of $\UUUU_0=\R^\nnn$.
		\renewcommand{\algorithmicrequire}{\textbf{Output:}}
			\Require $T:=[W_\nnn,W_{\nnn-1},\ldots,W_1]^{-1}$.
		\end{algorithmic}
	\end{algorithm}

		In Algorithm~\ref{alg1}, Proposition~\ref{prop:basis} in Appendix ensures that the columns of $[W_1,W_2,\ldots, W_{i-1}, FW_{i-1}]$ are independent vectors in the space $\UUUU_{\nnn-i}$, for each $i=2,\ldots, \nnn$.
		In case they already compose a basis of $\UUUU_{\nnn-i}$, then the matrix $V_i$ is defined by a null matrix.

		Now, we state the result.
		Let $k_i$, $i=1,\ldots,\nnn$, denote the number of columns of the matrix $W_i$, so that $W_i\in\R^{\nnn\times k_{i}}$.
		It is clear that $\sum_{i=1}^\nnn k_i = \nnn$ with $k_1 \le k_2\le \ldots \le k_\nnn$,
		and $k_\nnn$ is equal to the dimension of the controller output, i.e., $k_\nnn=\mmm$. This is because, by definition, the columns of $\{W_i\}_{i=1}^{\nnn-1}$ compose a basis of $\UUUU_1=\ker H$, whose dimension is $\nnn-\mmm$ since $H$ is of full row rank.
		Then, the following proposition shows that the transformation $T$ leads to a canonical form that enables to have both the state matrix and output matrix as integers.
		
		\begin{prop1}\label{prop:output}
			The following holds.
			\begin{enumerate}
				\item The transformation $T$ yields
				\begin{subequations}
					\begin{align}
						TFT^{-1} &= \begin{bmatrix}
							A_{1}&\Gamma_{\nnn-1} &0 &\cdots &0 \\
							A_{2}&0 &\Gamma_{{\nnn-2}} &\ddots &\vdots \\
							A_{3}&0 &0 &\ddots &0 \\
							\vdots&\vdots &\vdots &\ddots &\Gamma_{1} \\
							A_{\nnn}&0 &0 &\cdots &0
						\end{bmatrix}\label{eq:prop11}\\
						HT^{-1}&=\begin{bmatrix}
							HW_\nnn&0&0&\cdots&0
						\end{bmatrix} \label{eq:prop12}
					\end{align}
				\end{subequations}
				where $\Gamma_i=[I_{k_i}, 0]^\top\in\R^{{k_{i+1}}\times{k_i}}$ for $i=1,\ldots,\nnn-1$, and $[A_1^\top,\ldots,A_\nnn^{\top}]^\top = TFW_\nnn\in\R^{\nnn\times \mmm}$.
				Furthermore, the matrix $HW_\nnn\in\R^{\mmm\times \mmm}$ is invertible.
				\item With $R=FW_\nnn(HW_\nnn)^{-1}$ and $T_u = (HW_\nnn)^{-1}$, both the state matrix $T(F-RH)T^{-1}$ and the output matrix $T_uHT^{-1}$ consist of
				$\{0,1\}$, so that \eqref{eq:output_integer} holds.
				\hfill$\square$
			\end{enumerate}
		\end{prop1}
		
		{\it Proof:} 1)
		Recall that $T^{-1} = [W_\nnn,W_{\nnn-1},\ldots,W_1]$.
		The columns of $\{W_i\}_{i}^{\nnn-1}$ compose a basis of $\UUUU_1=\ker H$, so that $HW_i=0$ for $i\le \nnn-1$. Thus, we have \eqref{eq:prop12}.
		Since $H$ is of full row rank, $HW_\nnn\in\R^{\mmm\times\mmm}$ is invertible.
		Next, note that $$TFT^{-1} = T\begin{bmatrix}
			FW_\nnn&FW_{\nnn-1}&\cdots&FW_1
		\end{bmatrix}.$$
		From the definition $W_i:= [FW_{i-1},V_i]$, for $i\ge 2$, it is easy to verify that
		\begin{multline*}
			TFW_{i-1} = T\begin{bmatrix}
				FW_{i-1}&V_{i}
			\end{bmatrix}\Gamma_{{i-1}} = TW_i\Gamma_{{i-1}}\\
			= T\begin{bmatrix}
				W_\nnn & \cdots& W_i  &\cdots& W_1
			\end{bmatrix}\begin{bmatrix}
				0\\\vdots\\\Gamma_{{i-1}}\\\vdots\\0
			\end{bmatrix} = \begin{bmatrix}
				0\\\vdots\\\Gamma_{{i-1}}\\\vdots\\0
			\end{bmatrix}.
		\end{multline*}
		From this, we have \eqref{eq:prop11} proven.
		2) It is obvious from \eqref{eq:prop12} that $T_uHT^{-1}= [I_\mmm,0,\ldots,0]$. The proof is completed by
		\begin{align*}
			T(F-RH)T^{-1} &= TFT^{-1} -  (TR)(HT^{-1})\\
			&= TFT^{-1} - TFW_\nnn\begin{bmatrix}
				I_\mmm&0&\cdots&0
			\end{bmatrix}\\
			&=\begin{bmatrix}
				0&\Gamma_{\nnn-1} &0 &\cdots &0 \\
				0&0 &\Gamma_{{\nnn-2}} &\ddots &\vdots \\
				0&0 &0 &\ddots &0 \\
				\vdots&\vdots &\vdots &\ddots &\Gamma_{1} \\
				0&0 &0 &\cdots &0
			\end{bmatrix}
		\end{align*}
		because $[A_1^\top,\ldots,A_\nnn^{\top}]^\top = TFW_\nnn$.
		\hfill$\blacksquare$

		In case the controller \eqref{eq:controller} has been designed with rational matrices,
		the following corollary implies that, the converted system \eqref{eq:controller_transformed} can keep the input matrices $\{TG, TR\}$ as rational numbers as well, so that they can be kept in digital computers without truncation errors.
		\begin{cor}\label{cor:1}
			If $F$ and $H$ consist of rational numbers, then  $\{T,R,T_u\}$ can be found as rational matrices.
		\end{cor}
		
		{\it Proof:}
		In \eqref{eq:chain} and \eqref{eq:unobs}, each basis of $\UUUU_i$ can be found consisting of rational vectors, by virtue of linear algebra over rational numbers.
		It follows that all the matrices $\{W_i\}_{i=1}^{\nnn}$, and $\{T,R,T_u\}$ are found rational numbers.\hfill$\blacksquare$

		Thanks to Proposition~\ref{prop:output}, with $\{T,R,T_u\}$ found, the controller \eqref{eq:reenc} can be implemented over integers, as
		\begin{align}\label{eq:integer}
			\begin{split}	
				\ol z(t+1) &= T(F-RH)T^{-1}\ol z(t) \\
				&\quad  + \left\lceil\frac{TG}{\sss}\right\rfloor\left\lceil\frac{y(t)}{\rrr}\right\rfloor+ \left\lceil\frac{TR}{\sss}\right\rfloor\left\lceil\frac{\uuu(t)}{\rrr}\right\rfloor\\
				\ol u_z(t) &= T_uHT^{-1} \ol z(t)
			\end{split}		
		\end{align}
		with $\ol z(0) = \lceil(Tx_0)/(\rrr\sss)\rfloor$, where $\rrr>0$ is the quantization step size, $1/\sss \ge 1$ is a scale factor,
		and $\uuu(t)\in\R^\mmm$ is the output approximate to $u(t)$ of \eqref{eq:controller}, obtained by
		$$ \uuu (t) =\rrr\sss\cdot  T_u^{-1} \ol u_z(t).$$
		
		Clearly, the system \eqref{eq:integer} operates exploiting $(+,\times)$ only.
		And in terms of performance compared to \eqref{eq:reenc} or \eqref{eq:controller},
		it can be easily verified that the size of the error
		\begin{multline*}
			\rrr\sss\left(\left\lceil\frac{TG}{\sss}\right\rfloor\left\lceil\frac{y(t)}{\rrr}\right\rfloor+ \left\lceil\frac{TR}{\sss}\right\rfloor\left\lceil\frac{\uuu(t)}{\rrr}\right\rfloor\right)\\
			- (TGy(t) + TR\uuu(t))	
		\end{multline*}
		can be made arbitrarily small by increasing $1/\rrr$ and $1/\sss$, so the error between the outputs $u(t)$ of \eqref{eq:controller} and $\uuu(t)$ of \eqref{eq:integer}
		will also be arbitrarily small, under stability \cite{Kim23TAC}.
		In case the input matrices $TG$ and $TR$ are of rational numbers, the scale factor $1/\sss$ can be found as a rational number such that both $TG/\sss$ and $TR/\sss$ become integer matrices.
		In such case, in \eqref{eq:integer}, the rounding operations for $TG/\sss$ and $TR/\sss$ are dispensable.

		Finally, the benefit in terms of the required size of the plaintext space for implementing \eqref{eq:integer} is discussed.
		Thanks to the output matrix $T_u H T^{-1}$ as integers, there is no need of additionally multiplying the scale factor $1/\sss$ to convert it to integers (just as the input matrices in \eqref{eq:integer}).
		Then, while computing the output $\ol u_z(t)$ of \eqref{eq:integer} from the inputs, the scale factor $1/\sss$ multiplied only once with the input matrix.
		If the output matrix was also scaled by $1/\sss$, it would need to multiply the factor $1/\sss$ twice for computing $\ol u_z(t)$, as in \cite{Kim23TAC}.
		Considering that increasing the factor $1/\sss$ is required to reduce the performance error, eliminating the factor from the output matrix will let the plaintext space to cover the range of $\ol u_z(t)$ be reduced in this sense.

		Furthermore,
		in case the matrices can be kept as rational numbers as shown by Corollary~\ref{cor:1},
		their exact values can be kept by a certain choice of $1/\sss$ as a rational number, and there is no need of increasing $1/\sss$ than that.
		The reduction of the plaintext space size will imply the reduction of the size of encrypted numbers, which will lead to the reduction of storage use and computational burden of the encrypted controller.
		
		\begin{rem}
			Several results \cite{TeranishiTCNS,LeeTSMC} using re-encryption represent the controller to auto-regressive models, as
			\begin{equation}\label{eq:arx}
				u(t) = \sum_{i=1}^{\nnn}A_i u(t-i) + \sum_{i=1}^{\nnn}B_iy(t-i)
			\end{equation}
			where $\{A_i,B_i\}_{i=1}^{\nnn}$ are certain matrices or scalars.
			In terms of the computational burden, with $\{y(t-i)\}_{i=1}^{\nnn}$, the system \eqref{eq:arx} costs $\nnn$-times of $\mmm$-by-$\ppp$ matrix multiplication for each unit of time.
			Compared to this, the system \eqref{eq:controller_transformed} (and \eqref{eq:integer}) consumes two times of matrix multiplication by $TG\in\R^{\nnn\times \ppp}$ and $TR\in\R^{\nnn\times \mmm}$, for reflecting $\{y(t),\uuu(t)\}$.
			And, since the state matrix $T(F-RH)T^{-1}$ is an upper triangular matrix consisting of $\{0,1\}$ only, when stored as plaintexts, the state matrix multiplication can in fact be performed by simple shift operations.
			In this regard, the proposed method will be competitive, particularly when the input and output dimensions $\ppp$ and $\mmm$ are large.
		\end{rem}
		
		\subsection{Method for Intermittent Re-Encryption}\label{subsec:intermittent}
		
		We get back to the point that the controller \eqref{eq:reenc} can operate over integers, under the premise that the portion $u(t)$ is fed every time $t$ as an auxiliary input.
		This means, in terms of encryption, that the encrypted outcome of $u(t)$ has to be transmitted to the decryption device having secret key, decrypted, re-encrypted, and again transmitted back to the controller, within every sampling period.
		
		To increase the period of re-encryption $k$-times with some $k$, we first increase the update period for the state $x(t)$ of \eqref{eq:controller} by $k$-times;
		let the system \eqref{eq:controller} be replaced with
		\begin{subequations}\label{eq:intermittent}
			\begin{equation}\label{eq:intermittent_state}
				x(t+k) = F^k x(t) + G_k Y(t,k),\quad t=0,k,2k,\ldots
			\end{equation}
			in which we define
			\begin{align*}
				G_i &:= \begin{bmatrix}
					F^{i-1}G & F^{i-2}G& \cdots & G
				\end{bmatrix}\in\R^{\nnn \times i\ppp}\\
				Y(t,i) &:=\begin{bmatrix}
					y(t)\\
					y(t+1)\\
					\vdots\\
					y(t+i-1)
				\end{bmatrix}\in\R^{i\ppp },\quad\text{for}~~ i=0,1,\ldots,k
			\end{align*}
			so that the state $x(t)$ is updated every $k$-th time step.
			In case $i=0$,
			we simply consider $G_i$ and $Y(t,i)$ a null matrix and a null vector, respectively.
			Then,
			during each period of $k$-th time step,
			the output $u(t)$ can be obtained by
			\begin{equation}\label{eq:intermittent_output}
				u(t+i) = HF^i x(t) +HG_i Y(t,i),\quad i=0,1,\ldots,k-1
			\end{equation}
		\end{subequations}
		where $t=0,k,2k,\ldots$ is multiple of $k$.
		
		To implement
		\eqref{eq:intermittent} over $(\Z,+,\times)$, recall that its state matrix $F^k\in\R^{\nnn\times\nnn}$ needs to be converted to integers.
		Assuming that now re-encryption of $u(t)$ is available at $t=0,k,2k,\ldots$ only, we can re-write \eqref{eq:intermittent_state} as
		\begin{equation}\label{eq:reenc_intermittent}
			x(t+k) = (F^k-RH) x(t) + G_k Y(t,k) + Ru(t),
		\end{equation}
		where $R\in\R^{\nnn\times \mmm}$ will be a design parameter again.
		
		Recall that the design of $R$ is to have the state matrix $F^k-RH$ be transformed to integers, as
		$$T(F^k-RH)T^{-1}\in\Z^{\nnn\times \nnn}, $$
		and we have seen that such $R$ exists if the pair $(F^k,H)$ is observable.
		However,
		the pair $(F^k,H)$ with $k\ge 2$ may not be observable in general.
		For example,
		when
		$$F = \begin{bmatrix}
			0 & 1 \\ 0& 0
		\end{bmatrix}~~ \text{or} ~~ \begin{bmatrix}
			-1& 0 \\ 0& 1
		\end{bmatrix},\quad H=\begin{bmatrix}
			1&1
		\end{bmatrix},\quad k=2, $$
		then $(F,H)$ is observable, but $(F^k,H)$ is not observable.

		Nonetheless, we show that
		$R$ and $k$ can always be chosen such that $F^k-RH$ can be transformed to integers.
		We choose the period $k$ such that
		any two distinct eigenvalues of $F$ taken to the power of $k$ remain distinct; i.e.,
		\begin{multline}\label{eq:distinct}
			\text{if}~~\lambda_1\neq \lambda_2, ~~\det(F-\lambda_1I_\nnn)=0,\\ \text{and}~~\det(F-\lambda_2I_\nnn)=0,~~
			\text{then}~~\lambda_1^k\neq \lambda_2^k
		\end{multline}
		where $\det$ denotes the determinant of a matrix.
		Note that most positive integers for $k$ will satisfy\footnote{The condition \eqref{eq:distinct} is not satisfied only when the polynomial $\rho(s)=\det(F-s I_\nnn)$ has a factor $s^l -\alpha$ with some $l\ge2$ and $\alpha\neq 0$, with $k$ being a multiple of $l$.
			Choosing $k$ avoiding this, \eqref{eq:distinct} will be satisfied.} \eqref{eq:distinct}.
		Then, the following proposition shows that our claim is true, and a method for choosing $T$ and $R$ is described in the proof.

		\begin{prop}\label{prop:intermittent}
			If the pair $(F,H)$ is observable and the period $k$ is chosen such that \eqref{eq:distinct} holds, then there exist
			$T$ and $R$
			such that $T(F^k-RH)T^{-1}\in\Z^{\nnn\times\nnn}$.
		\end{prop}
		
		{\it Proof:} By use of Jordan canonical form,
		let an invertible matrix $T_0\in\R^{\nnn\times\nnn}$ be found such that
		\begin{align*}
			T_0FT_0^{-1} = \begin{bmatrix}
				F_1 & 0 \\
				0 & F_2	
			\end{bmatrix}~~\text{and}~~HT_0^{-1} = \begin{bmatrix}
				H_1 & H_2
			\end{bmatrix}
		\end{align*}
		where all the eigenvalues of $F_1\in\R^{\nnn_0\times\nnn_0}$, $\nnn_0\le \nnn$, are zero, and all the eigenvalues of $F_2\in\R^{(\nnn-\nnn_0)\times(\nnn-\nnn_0)}$ are non-zero.
		Note that the observability of $(F,H)$ implies the observability of $(F_2,H_2)$.
		We prove that $(F_2^k,H_2)$ is observable;
		choose an eigenvalue $\lambda\neq0$ of $F_2$ (so $\lambda^k$ is an eigenvalue of $F_2^k$), and suppose that there exists a vector $v\in\R^{\nnn-\nnn_0}$ such that  $(F_2^k-\lambda^kI_{\nnn-\nnn_0})v=0$ and $H_2v=0$.
		It follows that
		$$\prod_{l=0}^{k-1} \left(F_2-\lambda e^{\frac{2\pi i}{k}l}I_{\nnn-\nnn_0}\right)v=0.$$
		By the condition \eqref{eq:distinct}, when $l\neq 0$, the number $\lambda e^{\frac{2\pi i}{k}l}$ cannot be an eigenvalue of $F_2$, so that $F_2-\lambda e^{\frac{2\pi i}{k}l}I_{\nnn-\nnn_0}$ is invertible.
		This implies $(F_2-\lambda I_{\nnn-\nnn_0})v=0$. Since $(F_2,H_2)$ is observable, we have $v=0$, and therefore, $(F_2^k,H_2)$ is observable.
		Now, we can choose $R_2\in\R^{(\nnn-\nnn_0)\times \mmm}$ such that all eigenvalues of $F_2^k-R_2H_2$ are zero,
		and define
		\begin{equation*}
			R:= T_0^{-1}\begin{bmatrix}
				0\\ R_2
			\end{bmatrix}\in\R^{\nnn\times\mmm}
		\end{equation*}
		Finally, observe that
		$$F_{\sf conv}:=T_0(F^k- RH)T_0^{-1}  =\begin{bmatrix}
			F_1^k & 0 \\
			R_2H_1 & F_2^k - R_2H_2
		\end{bmatrix}.$$
		Since all the eigenvalues of $T_0(F^k- RH)T_0^{-1}$ are zero, the Jordan canonical form of
		$F_{\sf conv}$ is determined as
		$$T_1F_{\sf conv}T_1^{-1} = \begin{bmatrix}
			0&\gamma_1&0&\cdots&0\\
			0&0&\gamma_2&\ddots&\vdots\\
			0&0&0&\ddots&0\\
			\vdots&\vdots&\vdots&\ddots&\gamma_{\nnn_0-1}\\
			0&0&0&\cdots&0
		\end{bmatrix}\in\Z^{\nnn_0\times\nnn_0} $$
		with some $T_1\in\R^{\nnn\times\nnn}$,
		where $\gamma_i\in\{0,1\}$ for each $i=1,\ldots,\nnn_0-1$.
		With $T:= T_1T_0$, the proof is completed.
		\hfill$\blacksquare$
		
		In the proof of Proposition~\ref{prop:intermittent}, it can be seen that all the eigenvalues of the unobservable part of $F^k$ with respect to $H$ are zero, so that the unobservable part can also be transformed to integers (without help of the matrix $R$).

		Now, with the matrices $T$ and $R$ found, the remaining construction is straightforward.
		Analogous to \eqref{eq:integer},
		it is first transformed by $T$ and then scaled by $1/\rrr$ and $1/\sss$, as
		\begin{align}\label{eq:integer_intermittent}
			\begin{split}	
				\ol z(t+k) &= T(F^k-RH)T^{-1}\ol z(t) \\
				&\quad  + \left\lceil\frac{TG_k}{\sss}\right\rfloor\left\lceil\frac{Y(t,k)}{\rrr}\right\rfloor+ \left\lceil\frac{TR}{\sss}\right\rfloor\left\lceil\frac{\uuu(t)}{\rrr}\right\rfloor\\
				\ol u_z(t+i) &= \left\lceil\frac{HF^iT^{-1}}{\sss}\right\rfloor \ol z(t) + \left\lceil\frac{HG_i}{\sss^2}\right\rfloor\left\lceil\frac{Y(t,i)}{\rrr}\right\rfloor\\
				&\qquad\quad t=0,k,2k,\ldots,\quad i=0,1,\ldots,k-1
			\end{split}
		\end{align}
		where $\ol z(0) = \lceil(Tx_0)/(\rrr\sss)\rfloor$, and $\uuu(t)\in\R^\mmm$ is the output recovered from $\ol u_z(t)$, computed by
		$$\uuu(t) = \rrr\sss^2 \cdot \ol u_z(t). $$
		The matrices for computing the output vary with time $i$, but the change is of period $k$, so that the matrices can be computed (and encrypted), {a priori}.

		Thanks to Proposition~\ref{prop:intermittent}, the system \eqref{eq:integer_intermittent} operates over $(\Z,+,\times)$.
		Regarding the encrypted operation and re-encryption,
		note that the signal $\uuu(t)$, which should be re-encrypted from $\ol u_z(t)$, is supposed to be used for updating the state $\ol z(t+k)$, but only at $t=0,k,2k,\ldots$. Thus, this will allow a $k$-times longer period for the re-encryption than the previous methods.
		By increasing the parameters $1/\rrr$ and $1/\sss$, the effect of rounding errors and the performance error can be made arbitrarily small, under stability
		(see \cite[Proposition 6]{Kim23TAC} or \cite[Lemma 2]{Kim22ARC} for more details).

		\section{Conclusion}
		Two modified methods have been proposed for running dynamic systems based on addition and multiplication over integers, to apply homomorphic encryption to networked controllers.
		The first method has been presented to convert both the state and the output matrices to integers, simultaneously, which will reduce the conservatism in term of the use of scale factors used for the controller matrices.
		And, considering that the length of sampling period would be regarded as a time limit for each unit of re-encryption (which requires access to the decryption key), the second method introducing the intermittent re-encryption will relieve this issue in practice.
		The use of re-encryption can be replaced with bootstrapping techniques of fully homomorphic encryption, because they can also be used for dividing scale factors.
		Main issue when using bootstrapping is computational cost and the time required, so this can also be addressed by adjusting the period of bootstrapping, with the proposed method.

		\appendix
		The following is used as a technical lemma.
		For a matrix $W$, let ${\mathrm{im}}W$ denote the image space of $W$.		
		\begin{prop}\label{prop:basis}
			Consider \eqref{eq:chain} with \eqref{eq:unobs}.
			Let a matrix $[W_1,W_2,\ldots,W_{j}]$ have $\nnn$-rows, its columns are linearly independent, and it satisfies that
			\begin{align}\label{eq:alg1}
				\begin{split}
					{\mathrm{im}}\,W_j&\subseteq\UUUU_{\nnn-j}\\
					{\mathrm{im}}\,W_j \cap \UUUU_{\nnn-j+1}&=\{0\}	\\
					{\mathrm{im}}[W_1,W_2,\ldots,W_{j-1}] &\subseteq \UUUU_{\nnn-j+1}.
				\end{split}
			\end{align}
			Then, the columns of $W:=[W_1,W_2,\ldots, W_{j}, FW_{j}]$ are independent vectors belonging to $\UUUU_{\nnn-j-1}$.
		\end{prop}
		
		{\it Proof:} It is easy to show that
		\begin{multline}\label{eq:app}
			\text{if}\quad x\in\UUUU_{\nnn-j} ~~\text{and}~~ x\not\in\UUUU_{\nnn-j+1},\\
			\text{then}\quad Fx\in\UUUU_{\nnn-j-1} ~~\text{and}~~ Fx\not\in\UUUU_{\nnn-j}.
		\end{multline}
		Thus,
		${\textrm{im}}\,FW_j \subseteq\UUUU_{\nnn-j-1}$, and hence ${\textrm{im}}\,W \subseteq\UUUU_{\nnn-j-1}$.
		Now, we show the linear independence.
		Suppose that
		\begin{equation}\label{eq:app2}
			\begin{bmatrix}
				W_1&W_2&\cdots&W_j
			\end{bmatrix}\alpha  + FW_j\beta = 0 
		\end{equation}
		with some column vectors $\alpha$ and $\beta$.
		By \eqref{eq:app} again, note that ${\textrm{im}}\,FW_j \cap \UUUU_{\nnn-j}=\{0\}$.
		Since \eqref{eq:app2} implies $FW_j\beta \in {\textrm{im}}\,[W_1,\ldots,W_j]\subseteq \UUUU_{\nnn-j}$, it follows that $\beta=0$.
		Then, linear independence of the columns of  $[W_1,\ldots,W_j]$ ensures that  $\alpha =0$, and it  completes the proof.\hfill$\blacksquare$
		
		It is obvious that the matrices $\{W_i\}_{i=1}^{\nnn}$ determined in Algorithm~\ref{alg1} satisfy \eqref{eq:alg1}, for each $j=1,2,\ldots,\nnn-1$.

\end{document}